\documentclass[twocolumn,secnumarabic, amssymb,amsmath,
showpacs,aps]{revtex4-1}

\usepackage{graphicx,color,epsfig,rotate}
\usepackage{times}
\usepackage{color}
\usepackage{graphicx}
\usepackage{dcolumn}
\usepackage{amssymb}
\usepackage{amsmath}
\usepackage{amsfonts}
\usepackage{docs}%
\usepackage{bm}%
\usepackage{wrapfig}
\usepackage[colorlinks=true,linkcolor=blue,pagecolor=blue,filecolor=blue,menucolor=blue,urlcolor=blue,citecolor=blue,anchorcolor=blue]{hyperref}%
\usepackage{bm}
\usepackage{graphics}
\usepackage{graphicx}
\usepackage{color}
\usepackage{amssymb}
\usepackage{standalone}
\usepackage[toc,page]{appendix}
\usepackage{import}
\usepackage{enumerate}
\usepackage{sidecap}

\begin{document}

\title{ Supercurrent Reversal in Two-Dimensional Topological Insulators}
\author{Alexander Zyuzin}
\author{Mohammad Alidoust }
\author{Jelena Klinovaja}
\author{Daniel Loss}
\affiliation{Department of Physics,
University of Basel, Klingelbergstrasse 82, CH-4056 Basel,
Switzerland}



\begin{abstract} 
We calculate supercurrent across a two-dimensional topological insulator subjected to an external magnetic field. 
When the edge states of a narrow two-dimensional topological insulator are hybridized, an external magnetic field can close the hybridization gap, thus driving a quantum phase transition from insulator to semimetal states of the topological insulator. We find a sign reversal of the supercurrent at the quantum phase transition revealing intrinsic properties of topological insulators via Josephson effect.
\end{abstract}
\date{\today}

\maketitle

\section {Introduction}

Two-dimensional topological insulators (TIs) in the presence of time reversal symmetry are insulators in bulk while they possess gapless metallic edge states \cite{Pankratov, RevModPhys.82.3045}. These edge states are helical, i.e. they support counter-propagating modes with opposite spin projections, and are protected against elastic backscattering. 
One interesting phenomenon that may arise due to the rigid spin-momentum locking is 
topological superconductivity \cite{Volovik, FuKane, PhysRevB.78.195125, PhysRevLett.102.187001}. 
The topological superconductivity can play an important role in the realization of non-Abelian statistics needed for topological quantum computation \cite{RevModPhys.82.3045,RevModPhys.83.1057}.
To establish superconducting edge states in TIs, one can proximitize these states with a superconductor which results in the forming of Cooper pair wave functions in the TI (reviews on the proximity effect can be found in Refs. \onlinecite{Efetov_Rev,Buzdin_Rev}, for instance).

The edge states of a two-dimensional TI were experimentally investigated
in two different material classes ($\textrm{HgTe}/\textrm{HgCdTe}$ and $\textrm{InAs}/\textrm{GaSb}$) 
through imaging magnetic fields produced by corresponding edge currents and determining an effective edge resistance using a scanning superconducting quantum interference device (SQUID) \cite{ImageHgTE, PhysRevLett.113.026804}. 
In order to reveal the topological superconductivity in these edge channels, the current-voltage characteristics and Fraunhofer interference patterns in superconductor - two-dimensional TI - superconductor junctions were recently studied both theoretically \cite{Sim1, Sim2, Sim4, PhysRevLett.113.197001, PhysRevB.83.220511, PhysRevB.92.035428} and experimentally \cite{bib:Exp5, bib:Exp6, bib:Exp3, bib:Exp4}. 
In particular, the experimental study of the Fraunhofer interference patterns may allow one to argue for an edge-dominated contribution to the Josephson current although not conclusively \cite{bib:Exp3, bib:Exp4}. 

\begin{figure}[b]
\centering
     \includegraphics[height=5.7cm,width=80mm]{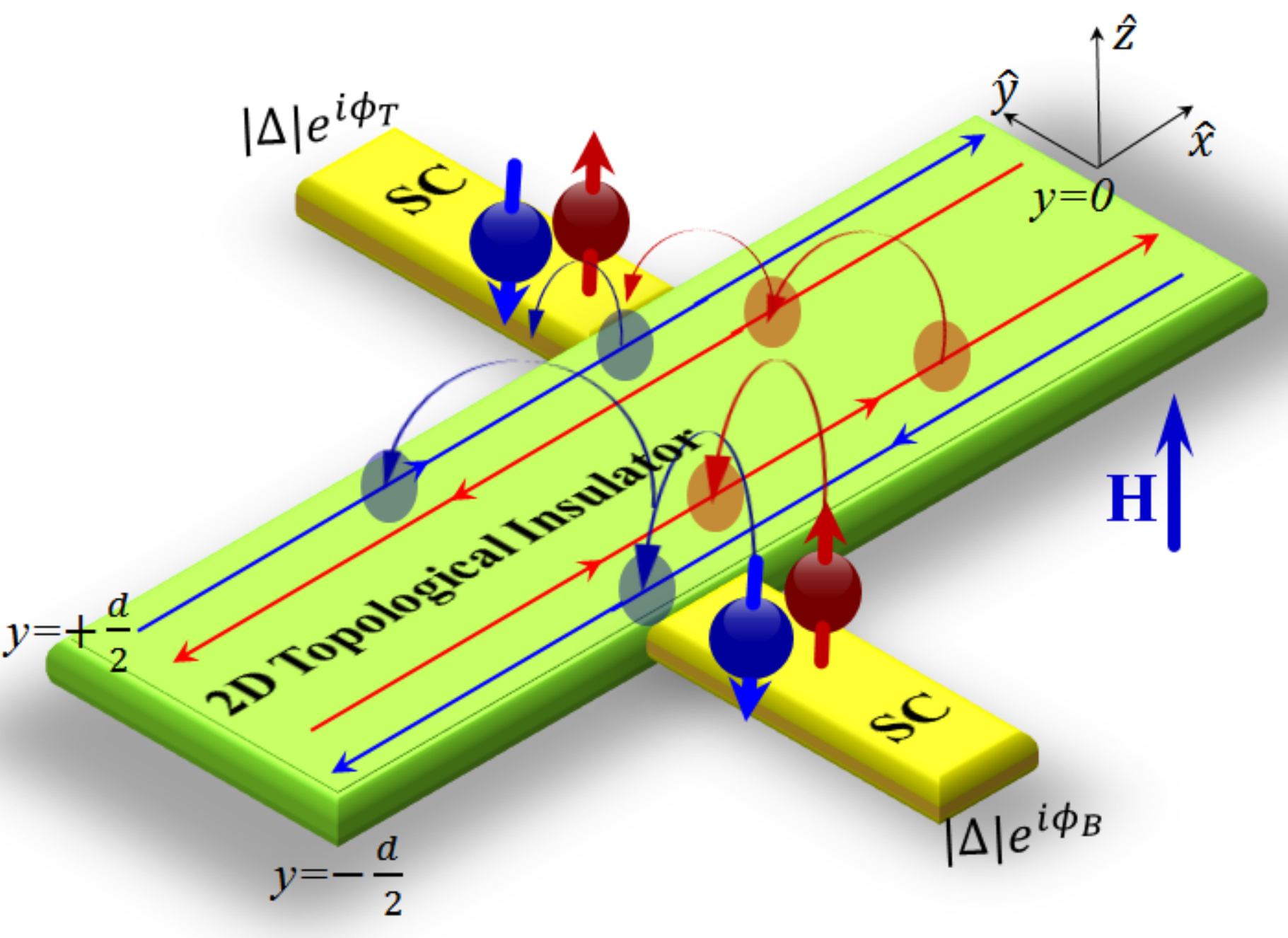}
     \caption{(Color online). Schematics of the point contact superconductor (SC) - 2D topological insulator - superconductor junction. The topological insulator strip has a width of $d$ and the junction plane resides in the $z=0$ plane so that the superconducting contacts are located at $\mathbf{ r}_{T/B}= (x_{T/B},\pm d/2, 0)$. 
The external magnetic field $\mathbf{ H} = (0,0,H)$ is oriented along $z$, normal to the plane of the TI. Top and bottom edge states (shown by parallel arrow lines) are hybridized while left and right edge channels are strongly gapped compared to the region of contact. 
The top and bottom superconducting electrodes have different macroscopic phases marked by $\phi_T$ and $\phi_B$, respectively. The typical tunneling process of Cooper pairs between the superconductors through the edge channels is also illustrated.  
          }
\label{pic1}
\end{figure}

An important practical challenge in Refs. \onlinecite{bib:Exp3, bib:Exp4} might be 
the penetration depth of helical edge states into the bulk material. According to these experiments, the penetration depth could be comparable to the width of the TI strip. Therefore, the edge states on the opposite sides of the strip may hybridize and produce a gap in the spectrum of edge states \cite{bib:BandCalc}. 
However, the externally applied magnetic field normal to the TI plane can close this gap and result in a phase transition from insulator to semimetal \cite{bib:Volovik_Klinkhamer, bib:BandCalc,bib:ZyuzinBurkov,KlinovajaDL-TIC}. 
Here, we demonstrate that the latter apparent disadvantage can turn to give evidence for the existence of edge-mode superconductivity in a point contact Josephson configuration as sketched in Fig. \ref{pic1}. 
We show that, by tuning the chemical potential, such a hybridization between the edges in the presence of external magnetic field results in a phase transition and reversal of a supercurrent flowing across the edge channels. 

In what follows, we first describe the model of the Josephson junction, discuss the properties of the two dimensional TI band structure, and then illustrate how the interplay of the magnetic field and hybridization between the edge states affects the supercurrent.

\section{Model of the Josephson junction}

The Josephson point contact configuration we consider  in this paper  is schematically shown in Fig. \ref{pic1}. We choose the plane of the two-dimensional (2D) TI to reside in the $z=0$ plane, the top and bottom edges of the TI to be parallel with the $\hat{x}$ axis. The two superconducting electrodes are connected to the TI via top and bottom edges at ${\bf r}_T=(x_T,+d/2,0)$ and ${\bf r}_B=(x_B,-d/2,0)$, respectively. An external magnetic field is applied normal to the junction plane along the $z$ direction, ${\bf H}=(0,0,H)$ (we will assume $H>0$ henceforth).
We adopt the Landau gauge for the vector potential $\mathbf{A}= (- y H,0,0)$ throughout our calculations. In what follows, we will be using $\hbar =1$ units.
Therefore, the system shown in Fig. \ref{pic1} can be described by the following Hamiltonian:
\begin{eqnarray}\label{1}
\mathcal{H}= \mathcal{H}_{\mathrm{SC}}+\mathcal{H}_{\mathrm{TI}}+\mathcal{H}_{\mathrm{Tun}},
\end{eqnarray}
where the first term is the usual BCS Hamiltonian describing superconducting leads
\begin{eqnarray}\label{1.1}\nonumber
\mathcal{H}_{\mathrm{SC}}&=&\sum_{j} \int_{V_{j}} d\mathbf{ r} \bigg\{\sum_{\alpha}\Phi^{\dag}_{j,\alpha}(\mathbf{ r})\bigg[\frac{1}{2m}(-i\nabla_{\mathbf{ r}} +e\mathbf{A}/c)^2\\
&-&\mu\bigg]\Phi_{j,\alpha}(\mathbf{ r})
+\bigg[\Delta_j(\mathbf{ r})\Phi_{j,\uparrow}^{\dag}(\mathbf{ r})\Phi_{j,\downarrow}^{\dag}(\mathbf{ r})+\mathrm{h.c.}\bigg]\bigg\},~
\end{eqnarray}
where $\Phi^{\dag}_{j,\alpha}(\mathbf{ r})$ is the electron creation operator in an SC lead, $j=T/B$ labels top/bottom SC lead, $\alpha= \uparrow, \downarrow$ stands for the spin quantum number, 
$\mu=p_F^2/2m$ and $m$ are chemical potential and electron mass, and $\Delta_{j}(\mathbf{ r})$ is the superconducting order parameter. The integration is performed over volume $V_{T/B}$ of the top/bottom SC. 
We consider magnetic field to be smaller than the first critical field of the SC so that orbital effects are weak and also neglect the Zeeman effect in the SC. 

The second term in Eq. (\ref{1}) stands for the Hamiltonian of the TI strip,
\begin{eqnarray}\label{2}\nonumber
\mathcal{H}_{\mathrm{TI}}&=& \sum_{j,j';\alpha,\alpha'}\int dx \Psi^{\dag}_{j,\alpha}(x)\bigg[v(-i\partial_x\tau^z_{jj'}-\kappa_H\delta_{jj'})\sigma_{\alpha\alpha'}^z \\
&+& t\tau^x_{jj'}\delta_{\alpha\alpha'}-\delta\mu \delta_{jj'}\delta_{\alpha\alpha'}\bigg]\Psi_{j',\alpha'}(x),~~
\end{eqnarray}
where 
$\Psi^{\dag}_{T/B,\alpha}(x)$ is the electron creation operator in the top/bottom edge states,
$v$ is the Fermi velocity, characterizing the edge 
dispersion, $t$ is the tunneling amplitude between top and bottom edge states ( 
which ensures electron spin and momentum conservations and we choose $t>0$ for concreteness),
$\sigma^{i}$ and $\tau^{i}$, $i=x,y,z$, are the Pauli matrices acting on the spin and the top/bottom edge of the TI pseudospin degrees of freedom, respectively. The position of chemical potential in the TI is measured from the charge neutrality point and is defined by $\delta\mu$. We have also defined a magnetic wave-vector
\begin{equation}
\kappa_H = d/ 2\ell^2 - g\mu_B H/2v,
\end{equation} 
containing contributions from the Aharonov-Bohm phase gradient (first term) and Zeeman coupling (second term), in which $\ell = \sqrt{c/eH}$ is the magnetic length, $g$ and $\mu_B$ are the electron g-factor and Bohr magneton, respectively.

The third term in Eq. (\ref{1}) describes the tunneling of electrons between superconducting leads and edge states of TI,
\begin{eqnarray}\label{3}\nonumber
\mathcal{H}_{\mathrm{Tun}}&=&\sum_{
   \alpha}
  \int dx \bigg[t_0(x)\Phi_{B,\alpha}^{\dag}(x,-d/2,0) \Psi_{B,\alpha}(x) \\
  &+&t_0(x)\Psi^{\dag}_{T,\alpha}(x)\Phi_{T,\alpha}(x,d/2,0)+ \mathrm{h.c.}\bigg],~~~
\end{eqnarray}
where $t_0(x)$ is the tunneling amplitude between the superconductor and the edge states. In our calculations we consider the geometry depicted in Fig. \ref{pic1} where the electron spin is conserved in the tunneling process between superconductor and TI whereas its momentum is not.

\begin{figure}[top]
     \includegraphics[width=80mm]{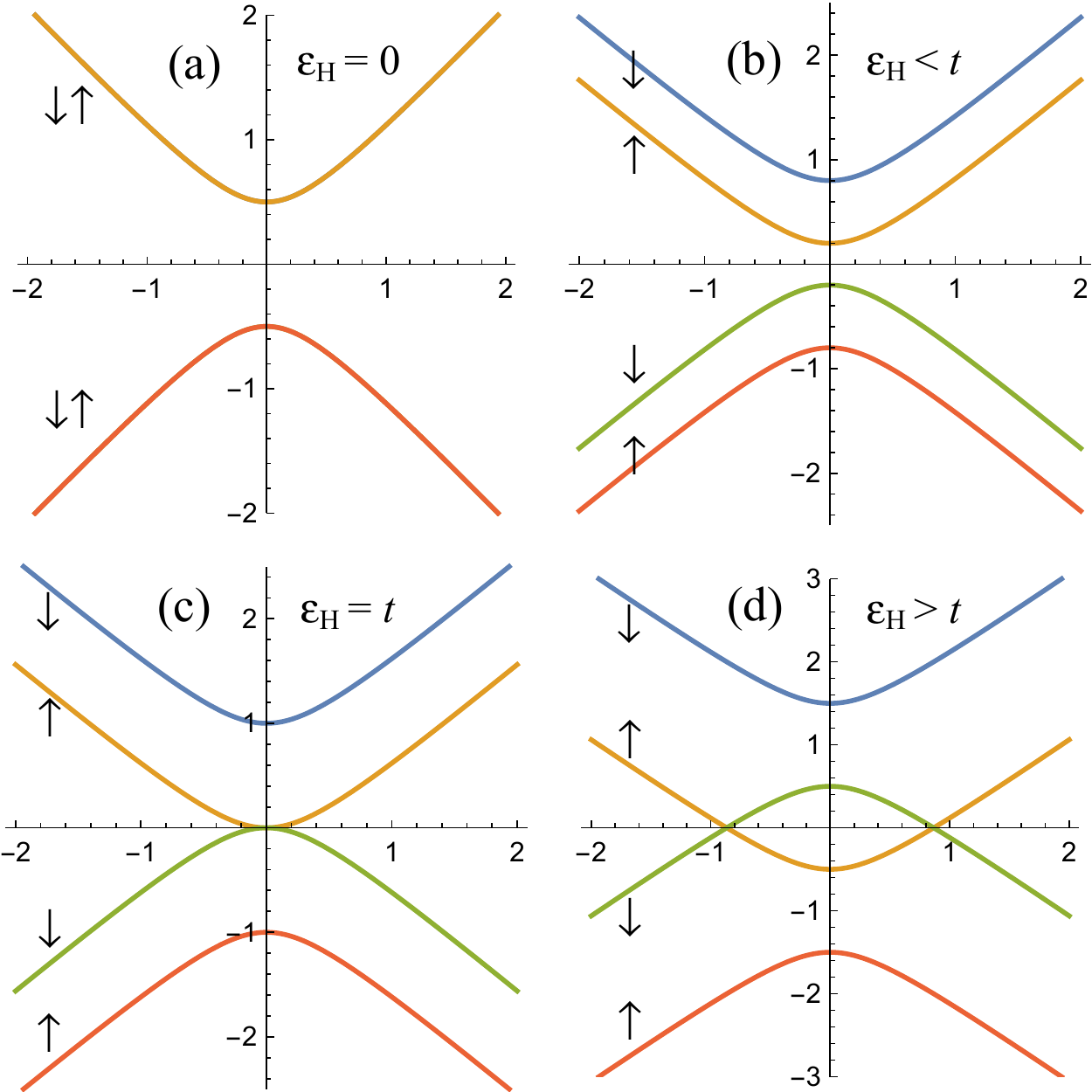}
     \caption{(Color online). Band structure, Eq. \ref{band}, of the 2D TI for different values of the magnetic energy $\varepsilon_H$ and the
     tunneling amplitude $t$. Arrows indicate the spin projection of an electron in the corresponding subband. (a)  The hybridization between the helical
     edge states with the same spin opens a gap of $2t$ at zero momentum. The subbands are doubly spin degenerate in the absence of an external magnetic field. (b) The external magnetic field removes the spin degeneracy and closes the gap. (c) Critical point $\varepsilon_H=t$ where two bands touch at $k=0$. (d) Subbands with opposite spins cross at two points when $\varepsilon_H > t$.
     }
\label{pic2}
\end{figure}

\section{Spectrum of hybridized edges states in magnetic field}

Now we discuss the properties of the TI edge band structure in the absence of superconducting leads. 
We shall return to the calculation of Josephson current later. Diagonalizing the Hamiltonian (\ref{2}) we obtain the edge states band dispersion,
\begin{equation}\label{band}
E_{s,s'}(k) =s\varepsilon_H +s'\sqrt{v^2k^2 + t^2},
\end{equation}
 where $s,s' = \pm 1$, $k$ denotes the momentum along the edge, and $\varepsilon_H=v |\kappa_H|$ is the ``magnetic energy". 
Figure \ref{pic2} illustrates the characteristics of the band structure for various values of $\varepsilon_H/t$ at the neutrality point $\delta \mu=0$. In the absence of an external magnetic field ($\varepsilon_H=0$) the spectrum consists of single valence and conduction bands split by a gap of size $2t$ at $k=0$, as shown in Fig. \ref{pic2}a. 
In this regime, the system is an insulator and each band is doubly degenerate in spin space. However, a
small magnetic field $\varepsilon_H<t$ lifts the spin degeneracy and splits both conduction and valence bands by $2\varepsilon_H$ at $k=0$ which is seen in Fig. \ref{pic2}b. 
At a critical field value, namely
$\varepsilon_H=t$, the lowest conduction and highest valence bands touch at a single point, $k=0$.
When $\varepsilon_H> t$ the system is gapless and the low energy part of the spectrum has two crossing points at $\pm k_0$, where 
$k_0= \sqrt{\varepsilon_H^2 - t^2}/v$, see Fig. \ref{pic2}c,d. 
Thus, the critical value of the magnetic energy, $\varepsilon_H=t$, defines 
a quantum phase transition between the semimetallic and insulating states of the TI narrow strip \cite{bib:Volovik_Klinkhamer, bib:BandCalc}. 
Below, we show that the insulator to semimetal phase transition can have nontrivial signatures in the Josephson current flowing through such a system.

\section{Evaluation of the Josephson current}

We assume that 
the tunneling amplitude $t_0$ between the TI edge states and superconducting electrodes is smaller than the one between the edge channels. This assumption allows us to neglect the proximity induced superconducting gap in the TI. 
The Josephson current $J$ in the lowest order in $t_0$
can be expressed as
\begin{eqnarray}\nonumber\label{JJ}
J&=&2eT\sum_{n,\alpha\alpha'} \int \prod_{i=1}^4dx_i \mathrm{Im} \bigg\{F_{B,\alpha\alpha'}(\omega_n; x_1, x_2) t_0(x_2 )\\\nonumber
&\times&G_{TB,\alpha'}(-\omega_n; x_3, x_2)t_0(x_3)F^{\dag}_{T,\alpha'\alpha}(\omega_n; x_3, x_4)\\
&\times&t_0(x_4)G_{TB,\alpha}(\omega_n; x_4, x_1)t_0(x_1)\bigg\},
\end{eqnarray} 
where $\omega_n=(2n+1)\pi T$ is the fermionic Matsubara frequency, $n\in \mathbb{Z}$, and $T$ is the temperature. More details on the derivation 
can be found in Appendix. To derive Eq. (\ref{JJ}) we have assumed that the coupling of superconducting leads via edge states along the perimeter of the TI strip is strongly suppressed in comparison to the direct tunneling across the TI. Such a condition can be experimentally achieved, for example, by inducing an insulating gap in the left/right edge states stronger than that of the region in the vicinity of the contacts.

The anomalous Green function inside the superconductor $F_{j,\alpha\alpha'}(\tau,\tau', \mathbf{ r},\mathbf{ r}')= \langle T_{\tau}\tilde{\Phi}_{j,\alpha}(\tau,\mathbf{ r})\tilde{\Phi}_{j,\alpha'}(\tau',\mathbf{ r}')\rangle$ (here the operator $\tilde{\Phi}_{j,\alpha}(\tau,\mathbf{ r})$ is in the Heisenberg representation)  in the frequency representation and quasiclassical approximation reads,
\begin{eqnarray}\label{F}
F_{j,\alpha\alpha'}(\omega_n; \mathbf{ r}, \mathbf{ r}') = i\sigma^y_{\alpha\alpha'}\int\frac{d\mathbf{ p}}{(2\pi)^3}\frac{|\Delta|e^{i\phi_{j} } e^{i\mathbf{ p}(\mathbf{ r}-\mathbf{ r}')}}{\omega_n^2+\xi_{\mathbf{ p}}^2+|\Delta|^2},~
\end{eqnarray}
where $\xi_{\mathbf{ p}} =\mathbf{ p}^2/2m -\mu$, the absolute value of the order parameter $|\Delta|$ is assumed to be fixed in both superconductors, and $\phi_{T/B}$ is the macroscopic phase of top/bottom lead. 

The hybridization of the helical edge states 
is described by the electron Green function $G_{TB,\alpha}(\tau,\tau';x, x')=-\langle T_{\tau}\tilde{\Psi}_{T,\alpha}(\tau,x)\tilde{\Psi}_{B,\alpha}^{\dag}(\tau',x')\rangle$,
which can be obtained from Eq. (\ref{2}) in the frequency representation as follows,
\begin{eqnarray}\label{G}
G_{TB,\uparrow/\downarrow}(\omega_n; x,x')=\int \frac{te^{ik(x-x')} dk/(2\pi)}{(i\omega_n+\delta\mu \pm\varepsilon_H)^2-v^2k^2-t^2}.~~~~~
\end{eqnarray}

It is worth noting that Eq. (\ref{JJ}) governs the direct tunneling process of two electrons with opposite spins across the hybridized edge states. 
\begin{figure}[top]
\centering
 \begin{tabular}{cc}
   \includegraphics[width=80mm]{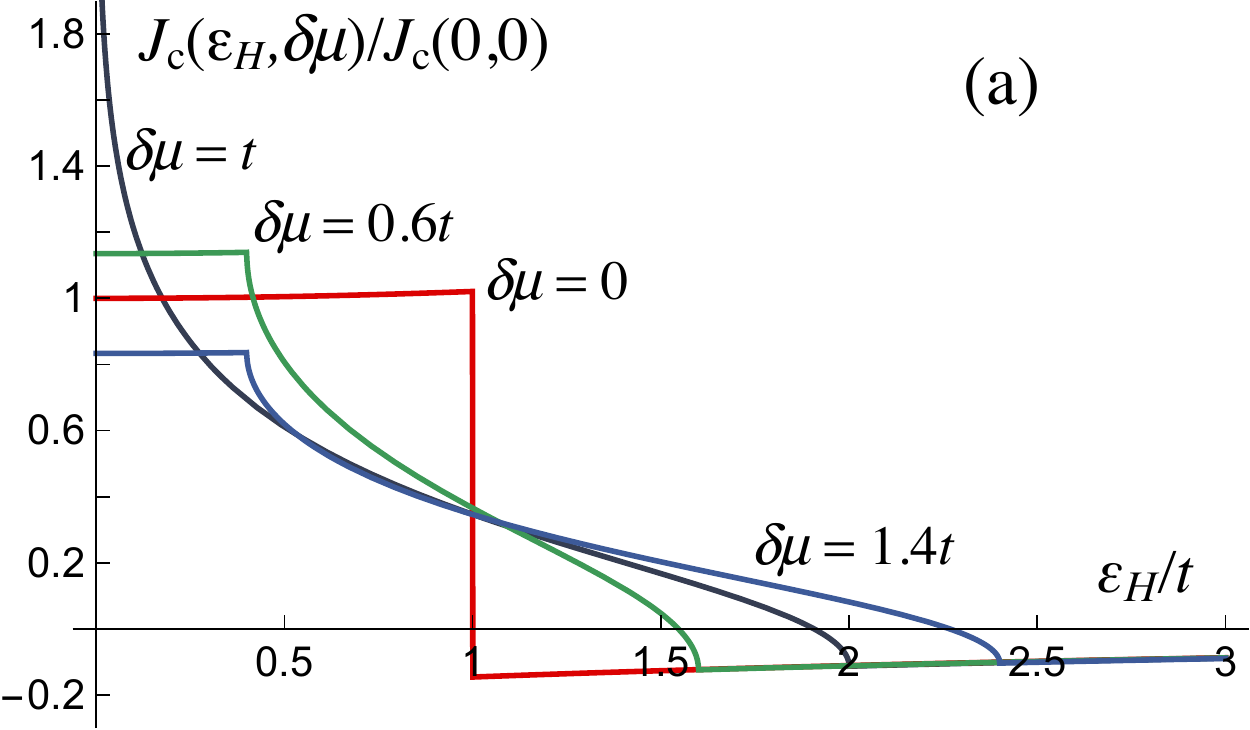}\\
          \includegraphics[width=80mm]{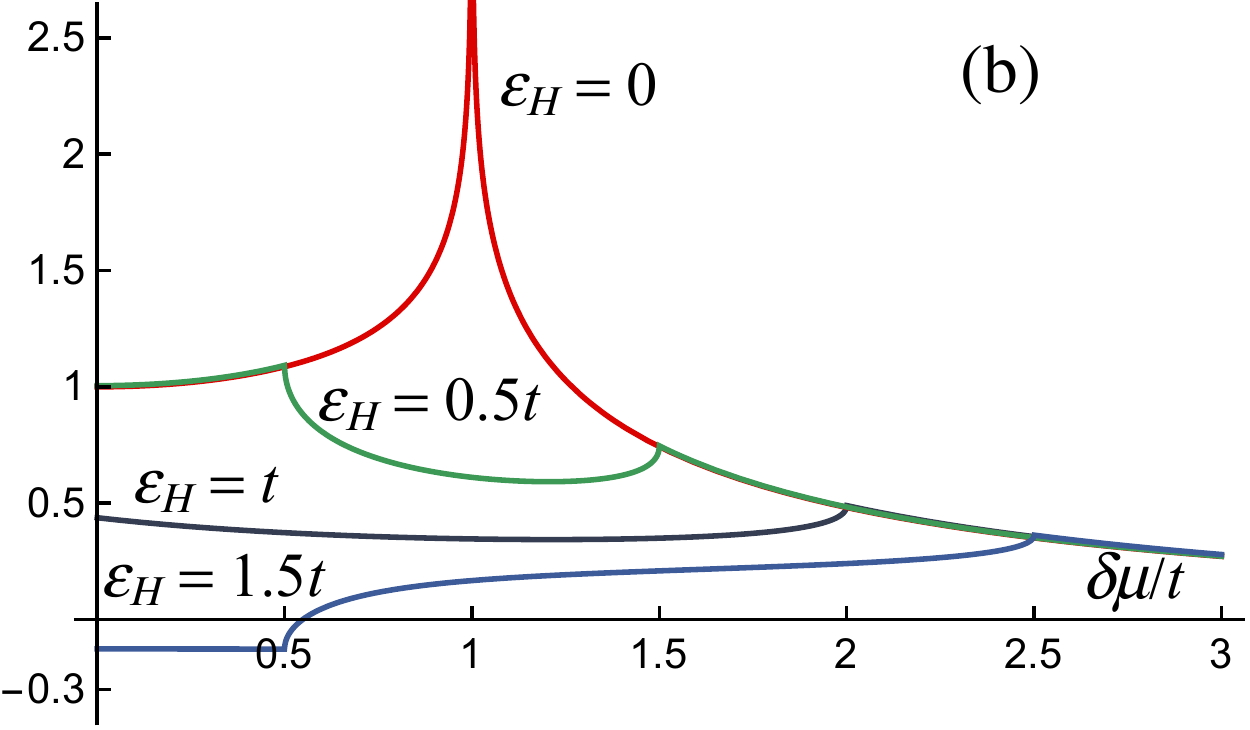}\\
  \end{tabular}
\caption{(Color online). Normalized critical supercurrent $J_c(\varepsilon_H,\delta\mu)/J_c(0,0)$ through the point contact configuration proposed where $|\Delta|/t =6$. 
(a). Critical current as a function of magnetic energy normalized by the tunneling amplitude at different values of chemical potential: $\delta \mu/t \in [0, 0.6,1,1.4]$. (b) Critical current as a function of chemical potential $\delta \mu$ normalized by the tunneling amplitude $t$ for different values of magnetic energy: $\varepsilon_H/t \in [0, 0.5,1,1.5]$.
}\label{pic3}
\end{figure}
We adopt a tunneling barrier model where the hopping amplitude $t_0(x) = t_0\delta(x-x_{T/B})$ is considered for top/bottom superconducting contacts in which $x_{T,B}\in [-L/2,L/2]$.
This assumption allows us to integrate the Josephson current, Eq. (\ref{JCpc}), over the lengths of top and bottom edge states involved in the supercurrent transport: $\langle  J \rangle = \int_{-L/2}^{L/2} dx_T dx_B J/L^2$. This double integral can be rewritten in terms of a double integral over the relative and center of mass coordinates.

Substituting the Green functions from Eqs. (\ref{F}) and (\ref{G}) into Eq. (\ref{JJ}) we obtain,
\begin{eqnarray}\nonumber\label{JCpc}
J&=& 4e \left[\pi^2 \nu_{\mathrm{TI}}\nu_{\mathrm{SC}} |\Delta| t t_0^2\right]^2 \sin(\phi_B-\phi_T)
\\\nonumber
&\times&T\sum_{n} \prod_s\frac{1}{\omega_n^2 + |\Delta|^2} \frac{e^{-|x_{B}-x_{T}|\sqrt{(\omega_n-is\delta\mu -i\varepsilon_H)^2+t^2}/v}}{[(\omega_n -is\delta\mu -i\varepsilon_H)^2+t^2]^{1/2}},\\
\end{eqnarray}
where $\nu_{\mathrm{SC}}=mp_F/2\pi^2$ is the density of states at the Fermi level per spin in the normal state of the superconductor leads, and $\nu_{\mathrm{TI}}=1/2\pi v$ is the density of states per spin and TI edge. 
We are interested in the zero temperature limit, $T = 0$, and transform the sum over frequencies into an integral, 
$T\sum_n \rightarrow\int \frac{d\omega}{2\pi}$. We emphasize that the integral in $\langle J \rangle$ behaves qualitatively different in the presence and absence of band crossing points in the spectrum, see Fig. \ref{pic2}. Indeed, we obtain a simple analytical expression for the Josephson current in the narrow junction limit $L \ll v/t$ of the form
\begin{eqnarray}\label{ResultPC}
\langle J\rangle=J_c(\varepsilon_H,\delta\mu) \sin(\phi_B-\phi_T), 
\end{eqnarray}
where the critical current across the contact at the charge neutrality point, $\delta\mu=0$, is given by,
\begin{eqnarray}\label{CC}
J_c(\varepsilon_H,0)&=&e |\Delta|t \left[\pi^2 \nu_{\mathrm{TI}}\nu_{\mathrm{SC}} t_0^2\right]^2 \\\nonumber
&\times&\left\{\frac{1}{|\Delta|+t+ \varepsilon_H} +\frac{\mathrm{sign}(t-\varepsilon_H)}{|\Delta|+|t-\varepsilon_H|}\right\}.
\end{eqnarray}
Focusing on the positive part of the TI spectrum, we see that the first term in $J_c(\varepsilon_H,0)$ contains contributions of the higher spin-down band at $\varepsilon_H<t$ (see Fig. \ref{pic2}). 
The magnetic field shifts the position of this band to higher energies. As a result, the tunneling between the leads via virtual states of this band is suppressed.  
The second term includes contribution from the low-energy spin-up band. The magnetic field in $\varepsilon_H<t$ region lowers the position of the band and increases the tunneling probability via its virtual states. This term changes sign at $\varepsilon_H=t$ and becomes negative at large magnetic fields $\varepsilon_H>t$. The external magnetic field inverts the conduction band curvature and flips the spin of the conduction band within $k\in [-k_0, k_0]$. The sign change in the supercurrent can be understood by noting the fact that the band-inversion permutes the order of tunneling of two electrons through the TI and is responsible for reversing the supercurrent direction. Further increase of the magnetic field suppresses the tunneling probability via this band. The negative coupling discussed here is similar to the Josephson junction with magnetic impurities studied in Refs. \onlinecite{Kulik, Bulaevskii, Spivak}.

To summarize, at the charge neutrality point, $\delta \mu=0$, and for small $\varepsilon_H < t$, an increase of the magnetic field increases the supercurrent $J_c(\varepsilon_H,0)$. When $\varepsilon_H = t$, the critical supercurrent changes sign. 
Further increase of the magnetic field decreases the absolute value of the current.
The critical current as a function of chemical potential $\delta\mu$ and magnetic energy $\varepsilon_H$ is shown in Fig. \ref{pic3}. We note that the supercurrent reversal can be observed as long as $\delta\mu$ is smaller than $|t-\varepsilon_H|$. Although we discussed our results within the zero temperature limit, a finite temperature only smears the singularities of Eq. (\ref{ResultPC}) in the energy interval $\sim T$ around the energy $|t\pm \delta \mu|$.

Having analyzed the supercurrent characteristics, we apply the condition of weak orbital regime 
$|\Delta|> v_Fd/ 2\ell^2 > \varepsilon_H\equiv v |\kappa_H|$, where the Fermi velocity in the superconducting leads, $v_F$, is usually larger than the Fermi velocity $v$ in the \textrm{TI}. Also in order to drive the
insulator-semimetal phase transition through the external magnetic field, we assume the tunneling element $t$ is much smaller than $|\Delta|$ so that the  condition $t=\varepsilon_H$ is satisfied. In this regime, the critical supercurrent strongly simplifies and is given by the following expression,
\begin{eqnarray}
J_{c}(\varepsilon_H,0)= 2e t (\pi^2 \nu_{\mathrm{TI}}\nu_{\mathrm{SC}} t_0^2)^2 [\Theta(t-\varepsilon_H)-\frac{t}{|\Delta|}].~~~~
\end{eqnarray}
We note that the amplitude of Josephson current through the semimetallic phase is therefore smaller than that of the insulator phase. It is also worth mentioning that the supercurrent reversal discussed above is not affected by the usual Fraunhofer response of critical supercurrent to an external magnetic field which can be observed in Josephson `wide' weak links \cite{abrikosov}. 

%
\section{Conclusions}

Let us now discuss experimental observability and feasibility of the experiment proposed in this paper to reveal the supercurrent reversal in 2D TIs.
We consider experimentally relevant parameter values for a 2D TI quantum well, $\mathrm{HgTe}/\mathrm{HgCdTe}$, adopted from the experiment reported in Ref. \onlinecite{bib:Exp3}. The separation between edge channels was reported equal to $d=400$$\mathrm{nm}$. Using band structure calculations, the hybridization energy associated with this value of $d$ was estimated about $t = 10$$\mu \mathrm{eV}$ \cite{bib:BandCalc}. This hybridization energy is of the same order as the proximity induced superconducting  gap in the edge states which was estimated by $\Delta_g\lesssim 20$$\mu \mathrm{eV}$ \cite{bib:Exp3}.
Using the edge mode velocity $v=5\times 10^7$$\mathrm{cm}/\mathrm{s}$ we find that the Zeeman contribution to $\varepsilon_H$ can be neglected compared to the contribution of Aharonov-Bohm phase gradient which results in a very small critical magnetic field $H=0.1$$\mathrm{mT}$. Therefore, a TI strip of width $d=270$$\mathrm{nm}$ will require an external magnetic field of order $H=1$$\mathrm{mT}$ to show the phase transition and supercurrent reversal discussed above \cite{bib:BandCalc}. These interesting effects can be experimentally verified by constructing the configuration shown in Fig. \ref{pic1} through the materials used in Refs. \onlinecite{bib:Exp3, bib:Exp4}. We are confident that such an experiment may pave the way to confirm the edge-mode superconductivity. 

\acknowledgements

We thank R. Tiwari and A. Yu. Zyuzin for helpful and interesting discussions and acknowledge support from the Swiss NF and NCCR QSIT.

\appendix

\begin{widetext}

\section{Josephson current}
Here we present more details of the tunneling supercurrent between two superconductors (SC) through a narrow strip of 2D topological insulator shown in Fig. 1 in the main text.
The tunneling between the SC leads and  edge states is treated perturbatively. The tunneling Hamiltonian in the interaction representation is given by
\begin{equation}
\mathcal{H}_{\textrm{Tun}}(\tau) = \sum_{\alpha}\int d\mathbf{ r} d\mathbf{ r}'~ [T_{\mathbf{ r},\mathbf{ r}'}\tilde{\Phi}^{\dag}_{\textrm{B},\alpha}(\tau,\mathbf{ r})\tilde{\Psi}_{\textrm{B},\alpha}(\tau,\mathbf{ r}') + T_{\mathbf{ r},\mathbf{ r}'}\tilde{\Psi}^{\dag}_{\textrm{T},\alpha}(\tau,\mathbf{ r})\tilde{\Phi}_{\textrm{T},\alpha}(\tau,\mathbf{ r}') + \textrm{h}.\textrm{c}.],
\end{equation}
where $\tilde{\Phi}^{\dag}_{j,\alpha}(\tau,\mathbf{ r})$ and $\tilde{\Psi}^{\dag}_{j,\alpha}(\tau,\mathbf{ r})$ are electron spin $\alpha =\uparrow,\downarrow$ creation operators in the top/bottom (denoted by index $j=T/B$) SC and top/bottom edge (denoted by the same index  $j=T/B$) of the TI, $T_{\mathbf{ r},\mathbf{ r}'}$ is the tunneling matrix element between TI and SC which is assumed to be a real quantity, $\tau$ is the imaginary time, and $\mathbf{ r}= (x,y,z)$. The imaginary time dependent current operator is defined by
\begin{equation}
\mathcal{I}(\tau) = ie \int d\mathbf{ r} d\mathbf{ r}' ~[T_{\mathbf{ r},\mathbf{ r}'}\tilde{\Psi}^{\dag}_{\textrm{B},\alpha}(\tau,\mathbf{ r})\tilde{\Phi}_{\textrm{B},\alpha}(\tau,\mathbf{ r}') - \textrm{h}.\textrm{c}.].
\end{equation}
Here the integrals over $\mathbf{ r}$ and $\mathbf{ r}'$ run over the edge states and the volume of SC, correspondingly. 
It is convenient to define a new operator as follows,
\begin{equation}
A_{j}(\tau)= \sum_{\alpha} \int d\mathbf{ r} d\mathbf{ r}' ~T_{\mathbf{ r},\mathbf{ r}'} \tilde{\Psi}^{\dag}_{j,\alpha}(\tau,\mathbf{ r}) \tilde{\Phi}_{j,\alpha}(\tau,\mathbf{ r}').
\end{equation}
Using this new operator, one can rewrite the current operator and the tunneling Hamiltonian,
\begin{equation}
 \mathcal{I}(\tau) = ie [A_{\textrm{B}}(\tau) - A_{\textrm{B}}^{\dag}(\tau)],
\end{equation}
\begin{equation}
 \mathcal{H}_{\textrm{Tun}}(\tau) = \sum_{j}[A_{j}(\tau) +A_{j}^{\dag}(\tau)].
\end{equation}

To find the Josephson current we take an average of the tunneling current operator in the lowest order of tunneling between the superconductor and TI. Therefore, we need to introduce an imaginary time evolution operator
\begin{equation}
S(\tau) = T_\tau \textrm{exp}\bigg\{-\int_0^{\tau} \mathcal{H}_{\textrm{Tun}}(\tau')d\tau' \bigg\}. 
\end{equation}
At zero bias voltage the single particle current between the SC electrodes vanishes. We also neglect the proximity induced superconducting minigap in the TI and set averages $\langle T_\tau \tilde{\Psi}_{i,\alpha}(\tau,\mathbf{ r}) \tilde{\Psi}_{j,\beta}(\tau',\mathbf{ r}') \rangle = \langle T_\tau \tilde{\Psi}^{\dag}_{i,\alpha}(\tau,\mathbf{ r}) \tilde{\Psi}^{\dag}_{j,\beta}(\tau',\mathbf{ r}') \rangle =0$ equal to zero. 
The current density in the fourth order of SC-TI tunneling matrix element reads,
\begin{equation}\label{crt}
J(\tau) =e \textrm{Im} \int_0^{1/T} d\tau_1 d\tau_2 d\tau_3 \langle T_{\tau} A_\textrm{B}(\tau)A_\textrm{B}(\tau_1)A^{\dag}_{\textrm{T}}(\tau_2)A^{\dag}_\textrm{T}(\tau_3)\rangle.
\end{equation}
The Green function in the SC leads is given by
\begin{eqnarray}
F_{j,\alpha\beta}(\tau-\tau'; \mathbf{ r},\mathbf{ r}') = \langle T_{\tau} \tilde{\Phi}_{j,\alpha}(\tau,\mathbf{ r}) \tilde{\Phi}_{j,\beta}(\tau',\mathbf{ r}') \rangle,\\
F^{\dag}_{j,\alpha\beta}(\tau-\tau'; \mathbf{ r},\mathbf{ r}') = \langle T_{\tau} \tilde{\Phi}^{\dag}_{j,\alpha}(\tau,\mathbf{ r}) \tilde{\Phi}^{\dag}_{j,\beta}(\tau',\mathbf{ r}') \rangle,
\end{eqnarray}
while the edge states of the TI are described by
\begin{eqnarray}
G_{ij,\alpha\beta}(\tau-\tau'; \mathbf{ r},\mathbf{ r}') = -\langle T_{\tau} \tilde{\Psi}_{i,\alpha}(\tau,\mathbf{ r}) \tilde{\Psi}_{j,\beta}^{\dag}(\tau',\mathbf{ r}')\rangle,\\
\bar{G}_{ij,\alpha\beta}(\tau-\tau'; \mathbf{ r},\mathbf{ r}') = -\langle T_{\tau} \tilde{\Psi}^{\dag}_{i,\alpha}(\tau,\mathbf{ r})\tilde{\Psi}_{j,\beta}(\tau',\mathbf{ r}') \rangle,
\end{eqnarray}
Performing a Fourier transformation to the Matsubara frequencies, one finds $F_{j,\alpha\beta}(\omega_n;\mathbf{ r},\mathbf{ r}') = T\sum_n e^{-i\omega_n\tau} F_{j,\alpha\beta}(\tau;\mathbf{ r},\mathbf{ r}')$, where $\omega_n = \pi T(2n+1)$, $n \in \mathbb{Z}$, and $T$ is the temperature. 
By substituting the Green functions introduced above into the expression (\ref{crt}), we arrive at the following expression for the Josephson current at zero bias voltage across the junction,
\begin{eqnarray}\nonumber
J &=& -2 e \textrm{Im} T\sum_{n,\alpha,\beta,\gamma,\delta}\int \prod_{j=1}^{8} d\mathbf{ r}_j F_{\textrm{B},\alpha\beta}(\omega_n;\mathbf{ r}_1,\mathbf{r}_2)T_{\mathbf{r}_2,\mathbf{r}_3}\bar{G}_{\textrm{BT},\beta\gamma}(\omega_n;\mathbf{r}_3,\mathbf{r}_4)T_{\mathbf{r}_4,\mathbf{r}_5}\\
&\times& F^{\dag}_{\textrm{T},\gamma\delta}(\omega_n;\mathbf{r}_5,\mathbf{r}_6)
T_{\mathbf{r}_6,\mathbf{r}_7}G_{\textrm{TB},\delta\alpha}(\omega_n;\mathbf{r}_7,\mathbf{r}_8)T_{\mathbf{r}_8,\mathbf{r}_1}.
\end{eqnarray}
This expression simplifies in the tunneling barrier model between the SCs and TI. Noting that the tunneling process takes place at the SC-TI boundaries we can write for top/bottom edges, $T_{\mathbf{r},\mathbf{r}'} = t_0(x)\delta(z)\delta(y\mp d/2)\delta(\mathbf{r}-\mathbf{r}')$ and obtain
\begin{eqnarray}\nonumber
J &=& -2 e \textrm{Im} T\sum_{n,\alpha,\beta,\gamma,\delta}\int \prod_{j=1}^{4} dx_j F_{\textrm{B},\alpha\beta}(\omega_n;x_1,x_2)t_{0}(x_2)\bar{G}_{\textrm{BT},\beta\gamma}(\omega_n;x_2,x_3)t_{0}(x_3)\\
&\times& F^{\dag}_{\textrm{T},\gamma\delta}(\omega_n;x_3,x_4)
t_{0}(x_4)G_{\textrm{TB},\delta\alpha}(\omega_n;x_4,x_1)t_{0}(x_1).
\end{eqnarray}
Here, we simplify our notation by writing: $F^{\dag}_{\textrm{T},\gamma\delta}(\omega_n;x_3,x_4) \equiv F^{\dag}_{\textrm{T},\gamma\delta}(\omega_n;(x_3, d/2, 0),(x_4, d/2, 0))$, $F_{\textrm{B},\alpha\beta}(\omega_n;x_1,x_2)\equiv F_{\textrm{B},\alpha\beta}(\omega_n;(x_1, -d/2, 0),(x_2, -d/2, 0))$, and $\bar{G}_{\textrm{BT},\beta\gamma}(\omega_n;(x_2, -d/2, 0),(x_3, d/2, 0))\equiv $ $G_{\textrm{TB},\delta\alpha}(\omega_n;x_4,x_1)\equiv G_{\textrm{TB},\delta\alpha}(\omega_n;(x_4, d/2, 0),(x_1, -d/2, 0))$.

\section{Green function in the topological insulator}
The Green function in the TI is written in the frequency and momentum representations, in the particle-hole, top-bottom edge pseudospin, and spin spaces as follows,
\begin{equation}
\underline{G}(\omega_n;k)=\begin{pmatrix}
\check{G}(\omega_n;k)& 0\\
0& \check{\bar{G}}(\omega_n;k)\\
\end{pmatrix};~~ \check{G}(\omega;k)=\begin{pmatrix}
\hat{G}_{TT}(\omega_n;k)& \hat{G}_{TB}(\omega_n;k)\\
\hat{G}_{BT}(\omega_n;k)& \hat{G}_{BB}(\omega_n;k)\\
\end{pmatrix},
\end{equation}
where
\begin{equation}
\hat{G}_{ij}(\omega_n;k)=\begin{pmatrix}
G_{ij,\uparrow\uparrow}(\omega_n;k)& G_{ij,\uparrow\downarrow}(\omega_n;k)\\
G_{ij,\downarrow\uparrow}(\omega_n;k)& G_{ij,\downarrow\downarrow}(\omega_n;k)\\
\end{pmatrix}.
\end{equation}
These functions satisfy Eq. (3) in the main text.
It is apparent that the Green function in the TI is spin-diagonal. In our calculations, we need top-bottom edge pseudospin off-diagonal components of the Green function which is defined by
\begin{eqnarray}
G_{\textrm{TB},\uparrow\uparrow/\downarrow\downarrow}(i\omega_n;k)&\equiv& G_{\textrm{TB},\uparrow/\downarrow}(i\omega_n;k)=G_{\textrm{BT},\uparrow/\downarrow}(i\omega_n;k) = \frac{t}{(i\omega_n+\delta\mu \pm \varepsilon_H)^2-v^2k^2-t^2}\\
\bar{G}_{\textrm{TB},\uparrow\uparrow/\downarrow\downarrow}(i\omega_n;k)&\equiv&\bar{G}_{\textrm{TB},\uparrow/\downarrow}(i\omega_n;k)=\bar{G}_{\textrm{BT},\uparrow/\downarrow}(i\omega_n;k) = \frac{-t}{(i\omega_n-\delta\mu\mp \varepsilon_H)^2-v^2k^2-t^2}.
\end{eqnarray}

\end{widetext}

\bibliography{SNS}

\end{document}